\documentclass[twocolumn,english,aps,groupaddress,prx]{revtex4-2}

\usepackage[colorlinks=true,allcolors=blue]{hyperref}
\usepackage{array}
\usepackage[latin9]{inputenc}
\usepackage{amsmath}
\usepackage{amssymb}
\usepackage{graphicx}
\usepackage{babel}
\usepackage{mathrsfs}
\usepackage{amsfonts}
\usepackage{epstopdf}
\usepackage{multirow}
\usepackage{color}
\usepackage{natbib}
\usepackage{bm}
\usepackage{amsthm}
\usepackage{tabularx}

\newcommand{\ri}{\mathrm{i}}

\usepackage{multirow}
\usepackage{pdfpages}
\usepackage{xcolor}
\usepackage{lipsum}
\usepackage{etoolbox}
\usepackage{float}
\usepackage{siunitx}
\usepackage{soul}
\usepackage{wasysym}
\usepackage[normalem]{ulem} 
\sethlcolor{yellow!35}
 
\makeatletter
\patchcmd{\@outputpage@head}{\@ifx{\LS@rot\@undefined}{}{\LS@rot}}{}{}{}
\makeatother

\begin{document}
\title{Tricritical Kibble--Zurek scaling in Rydberg atom ladders}
\date{\today}
\author{Hanteng Wang}
\affiliation{Institute for Advanced Study, Tsinghua University, Beijing, 100084, China}
\author{Xingyu Li}
\affiliation{Institute for Advanced Study, Tsinghua University, Beijing, 100084, China}
\author{Chengshu Li}
\email{chengshu@mail.tsinghua.edu.cn}
\affiliation{Institute for Advanced Study, Tsinghua University, Beijing, 100084, China}

\begin{abstract}
The Kibble--Zurek (KZ) mechanism has been extensively studied in various second-order phase transitions, yet the case of tricriticality---the point where second-order phase transition lines terminate---remains experimentally elusive. Here, we theoretically propose probing KZ scaling at tricritical points using Rydberg atom arrays arranged as two- and three-leg ladders, which realize the tricritical Ising and tricritical Potts universality classes. By slowly ramping the Rabi frequency and detuning, we extract two relevant tricritical exponents, $\nu$ and $\nu'$, both via conventional paths from the disordered to the ordered phase and via ``tangential'' paths confined entirely within the disordered phase. At faster speeds, ramping dynamics go beyond the standard KZ paradigm: data collapse analysis using the parent critical exponents (rather than the tricritical ones) reveals renormalization group flows toward the adjacent second-order critical line, and we identify it as a dynamical analog of Zamolodchikov's $c$-theorem. Our protocol is readily implementable on existing Rydberg quantum simulators. This provides a direct route to measuring distinct tricritical exponents which can reveal an emergent spacetime supersymmetry constraint $1/\nu - 1/\nu' = 1$. Moreover, this work deepens our theoretical understanding and opens new avenues for exploring beyond-KZ quantum dynamics with rich renormalization group structure.
\end{abstract}

\maketitle
The Kibble--Zurek (KZ) mechanism concerns dynamical scaling behavior near a critical point \cite{kibble1976topology,zurek1985cosmological}. Originally proposed in the context of classical phase transitions, its quantum counterpart has been extensively explored both within \cite{zurek2005dynamics,dziarmaga2005dynamics,polkovnikov2005universal,polkovnikov2011colloquium,Sandvik2011,kolodrubetz2012nonequilibrium,Sondhi2012,kolodrubetz2012nonequilibrium,del2014universality,rossini2021coherent} and beyond \cite{Bermudez2009Topology,bermudez2010dynamical,ChenInsulatorKZ2020,Nie2024topo,shu2025equilibration,Deng2025} the Landau--Ginzburg--Wilson paradigm.
With recent advances in quantum simulators, numerous experimental studies on artificial quantum platforms, such as superconducting qubits \cite{king2022coherent,king2023quantum}, ion traps \cite{ulm2013observation,pyka2013topological,li2023probing}, and Rydberg atom arrays \cite{keesling2019quantum,ebadi2021quantum,manovitz2025quantum,zhang2025near}, have emerged. The KZ ramping technique is particularly well-suited for probing critical phenomena for two complementary reasons. First, quantum simulators allow for versatile dynamical control and site-resolving measurements. Second, dynamical probing bypasses the challenge of adiabatic preparation of a gapless critical state \cite{fang2024probing}, where only a finite coherence time is available on these platforms.

At the intersection of first- and second-order phase transitions, a particular phenomenon known as tricriticality occurs. The tricritical point holds theoretical and practical importance. For instance, spacetime supersymmetry, originally proposed in high-energy physics, can emerge in the low-energy regime of various models \cite{grover2014emergent,jian2015susy,jian2017susy,li2018numerical,Li2020TCI,zeng2024nonequilibrium}, with one of the most experimentally promising cases being the $1+1$-dimensional tricritical Ising model \cite{Friedan1985,Zamolodchikov1986multicritical,li2024supersymmetry,cheng2025schwinger}. Tricritical points also play an important role in quantum annealing protocols, where ramping near these points can accelerate optimization processes \cite{wang2022mbl,zhang2024cyclic,zhang2025computational}. Furthermore, the proximity to first-order phase transitions raises questions about the validity of standard KZ scaling \cite{suzuki2024topological}, suggesting that new phenomena may emerge in this context. Despite its importance, no experimentally feasible protocol of tricritical KZ effect has been proposed~\cite{Mukherjee2007XY,divakaran2009defect,deng2009anomalous,mukherjee2010adiabatic,patra2011non}.

\begin{figure}[t!]
    \begin{center}  
    \includegraphics[width=0.9\columnwidth]{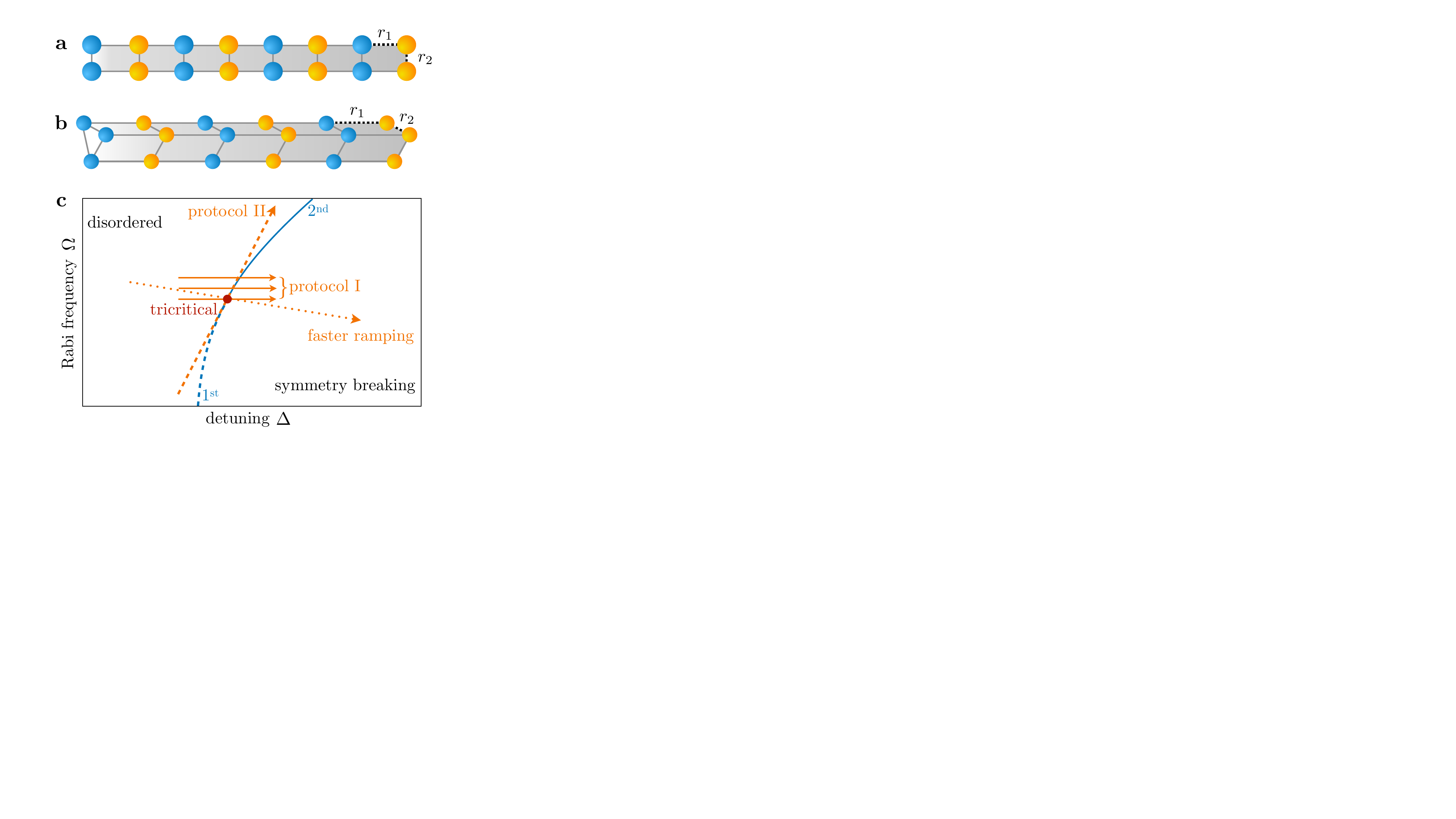}
    \caption{\textbf{Rydberg ladders and Kibble-Zurek protocols.} \textbf{a, b} Two- and three-leg ladders that exhibit tricritical Ising and Potts points, respectively. Atoms of different colors correspond to different atomic species. The geometry is fixed (i.e., $r_{1,2}$), while the Rabi frequency $\Omega$ and detuning $\Delta$ can be tuned. \textbf{c} The schematic phase diagram and ramping protocols. The solid (dashed) blue line denotes the second (first) order Ising/Potts transition, and a tricritical point lies in between. In Protocol I, we keep $\Omega$ fixed and ramp $\Delta$ from the disordered phase to the symmetry breaking phase, both across and near the tricritical point. In Protocol II, we consider a tangential ramping direction. The orange dot arrow refers to ramping of higher speeds, where a dynamical analog to Zamolodchikov's $c$-theorem is proposed.}
    \label{Fig:setup}
        \end{center}
\end{figure}

Here, we propose an experimentally accessible model based on Rydberg atom ladders \cite{Sarkar2023ladder,zhang2025ladder,kerschbaumer2024ladder,Chepiga2025ladder}, as illustrated in Fig.~\ref{Fig:setup}a and b. Figure \ref{Fig:setup}c presents the ground-state phase diagram, controlled by the Rabi frequency and detuning, with the key feature being the tricritical point connecting 1st- and 2nd-order (Ising/Potts) transition lines. The tricritical Ising (TCI) and tricritical Potts (TCP) transitions \cite{Nienhuis1979TCP} are realized respectively in the two- and three-leg ladders. From the perspective of renormalization group (RG) theory, these tricritical fixed points are characterized by two symmetry-preserving relevant operators \cite{DiFrancescoCFTBook,CardyBook}, introducing rich RG flow behaviors. These additional operators significantly impact the KZ dynamics, in contrast to the more widely studied critical points, which feature only a single symmetry-preserving relevant operator.

Furthermore, since the tricriticality only occurs at a single point in the phase diagram, generically (i.e. with non-vanishing deviation) only intermediate-distance physics is controlled by the tricritical point, with long-distance physics exhibiting more conventional behavior. Indeed, this is a concrete consequence of the celebrated Zamolodchikov's $c$-theorem \cite{zamolodchikov1986irreversibility}. Measuring tricritical exponents is therefore challenging in conventional materials, but this very task fits well with the capability of site-resolved imaging in quantum simulators, using dynamical finite-size scaling~\cite{Sandvik2011,liu2014dynamic,kolodrubetz2012nonequilibrium,rossini2021coherent,king2023quantum,zhang2025near}. We capitalize on this through the ramping Protocols I and II, as shown in Fig.~\ref{Fig:setup}c. Protocol I is of immediate experimental relevance. The conceptual novelty of Protocol II lies in the fact that, throughout the ramping, the system parameters are in the disordered phase, which is beyond the framework relating KZ to domain wall formation. These protocols provide a direct route to measuring distinct tricritical exponents, which can reveal an emergent spacetime supersymmetry constraint in the two-leg case. Moreover, by exploring a much wider range of ramping speeds, the results reveal intriguing crossover behavior from tricriticality to its parent second-order criticality, where a dynamical analog of Zamolodchikov's $c$-theorem is proposed.\\

\noindent {\large \textbf{Results}}\\
\noindent \textbf{Rydberg models and equilibrium phase diagrams} \\
\noindent The microscopic models, two- and three-leg Rydberg atom ladders, are modified from a previous model \cite{li2024supersymmetry}. In both the original and the current models, the system consists of two species of atoms \cite{Zhan2017PRL,Zhan2022PRL,Hannes2022PRX,anand2024dual}, assembled in an alternating pattern along the ladder, see Fig.~\ref{Fig:setup}a, b. Each atom is modeled by a two-state system, with the ground state $|g\rangle$ and the Rydberg state $|r\rangle$. A laser couples both states and determines the Rabi frequency $\Omega$ and detuning $\Delta$ through the laser strength and frequency, respectively. The two species are chosen such that Rydberg atoms of the same species repel, while those of different species attract. The Hamiltonian reads
\begin{equation}
H=\sum_{i}\left(\frac{\Omega}{2}\sigma^x_i-\Delta n _i\right)+\sum_{ij}\frac{C_{XY}}{r_{ij}^6}n_i n_j, \label{eq:Hamiltonian}
\end{equation}
where $X,Y$ denote the species of the atoms $i$ and $j$, and $r_{ij}$ is the inter-atom distance. $n=|r\rangle\langle r|$ is a projector onto the Rydberg state or, equivalently, the Rydberg number operator. $\sigma_x=|g\rangle\langle r|+\mathrm{h.c.}$ couples the ground state and the Rydberg state. $r_2$ is chosen to be reasonably small, so that only one atom within a rung can be excited to the Rydberg state due to Rydberg blockade.

In the original model, one fixes $r_2$ and $\Omega$, and tunes two parameters, namely $r_1$ and $\Delta$. In the current work, we consider instead tuning $\Omega$ and $\Delta$, keeping the geometry fixed throughout, see Fig.~\ref{Fig:setup}c. In fact, the phase diagram of the original model has the same structure as in Fig.~\ref{Fig:setup}c (with $\Omega$ replaced by $r_1$) thanks to universality (see also the numerical phase diagram in Supplementary Note 1). However, the new setup has the virtue that both $\Omega$ and $\Delta$ are determined by the laser, and are experimentally readily tuned in a dynamical fashion. In contrast, the lattice geometry is unchangeable during the dynamical evolution. Indeed, the atoms are arranged in a desired geometry by using optical tweezers. During the evolution, with atoms excited to Rydberg states, the tweezers could easily ionize the Rydberg atoms, causing significant atom loss. The upshot is that, in the original model, within each experimental run, one can only move in the horizontal (i.e. detuning) direction in the phase diagram. In the current setup, on the other hand, one is able to move along any path in the phase diagram.

The ladder models have a $\mathbb{Z}_{2}$ ($S_3$) symmetry that permutes the two (three) legs. For small $\Delta$, the atoms remain in the ground state, and the system is in the disordered phase. For large $\Delta$, the atoms prefer to be excited to the Rydberg state. However, with Rydberg blockade in action, the best one can do is to excite one of the two (three) legs. This breaks the symmetry to $\varnothing$ (reflective $\mathbb{Z}_2$), and hence we expect the transition to be Ising (three-state Potts) type. For small $\Omega$, the Hamiltonian is dominated by diagonal terms in the $Z$ basis, and the transition tends to be first-order. For large $\Omega$, the quantum fluctuation dominates, and the transition tends to be second-order. In between, TCI (TCP) points emerge, with distinct critical exponents compared to the Ising (Potts) case. All these predictions are corroborated by numerical simulations using density matrix renormalization group (DMRG)~\cite{White1992,MPS_2011}, see Methods for details. With the model and the phase diagram at hand, we next turn to dynamical evolutions. In the main text, we focus on the two-leg ladder. For the three-leg case, results repeat almost verbatim and we postpone them to the Supplementary Note 2. \\

\noindent \textbf{KZ scalings across and near TCI point} \\
\noindent To begin with, we consider ramping from the disordered phase to the ordered phase, similar to the conventional KZ protocol. This is shown in Fig.~\ref{Fig:setup}c as Protocol I, where the ramping parameter is given by $g = \Delta - \Delta_\mathrm{TCI}$. The ground-state correlation length diverges as $\xi_\mathrm{GS} \sim |g|^{-\nu}$, where $\nu$ is the critical exponent. Since the correlation length $\xi$ itself has length dimension, i.e., $[\xi] = -1$, the dimension of the ramping parameter is $[g] = 1/\nu$. Next, we define the dynamical process: ramping $g$ with speed $s$, i.e., $g = st$, with $t = 0$ set as the time passing the phase boundary. By this, one can recover the celebrated KZ scaling of the correlation length, $\xi_\mathrm{KZ} \sim s^{-\mu}$, by noting the dimension of speed: $[s] = [g] - [t] = 1/\nu + z \equiv 1/\mu$, where $z$ is the dynamical critical exponent.

In the quantum simulators we consider, however, the systems are finite-sized, meaning that the correlation length is limited at the critical point rather than diverging. This leads to a deviation from the $\xi_\mathrm{KZ} \propto s^{-\mu}$ scaling. Fortunately, as in the equilibrium case, finite-size scaling also applies to the KZ mechanism, namely $\xi_\mathrm{KZ}/L = \mathcal{F}(s^\mu L)$~\cite{Sandvik2011,liu2014dynamic,kolodrubetz2012nonequilibrium,rossini2021coherent,king2023quantum,zhang2025near}, where $L$ is the linear system size (e.g., the length of the ladder in our quasi-1D geometry). However, extracting the length scale $\xi_\mathrm{KZ}$ from the two-point correlation function introduces ambiguity, and this long-distance feature can easily be affected by noise \cite{zhang2025near}. Therefore, we choose the Binder ratio $U$ to characterize the dynamics. The main advantages of $U$ are twofold. First, it is dimensionless and does not scale by itself, thus eliminating one additional scaling parameter to be determined. Second, unlike the correlation length, it is a number defined without the need of fitting. Specifically, the time-dependent Binder ratio is defined through the order parameter $M$ as
\begin{equation}
U(t)=1-\frac{\langle \psi(t)| M^4 | \psi(t) \rangle}{3\langle \psi(t)| M^2 |\psi(t)\rangle^2}.
\end{equation}
The order parameter is defined as $M = \sum_i n_{ia} - n_{ib}$, where $a$ and $b$ denote the atoms within each rung, and $|\psi(t)\rangle$ is the wavefunction evolved by the Hamiltonian (\ref{eq:Hamiltonian}) up to time $t$. This quantity can be readily measured in Rydberg platforms using site-resolved \emph{in situ} imaging.

\begin{figure*}[t!]
    \begin{center}  
    \includegraphics[width=\textwidth]{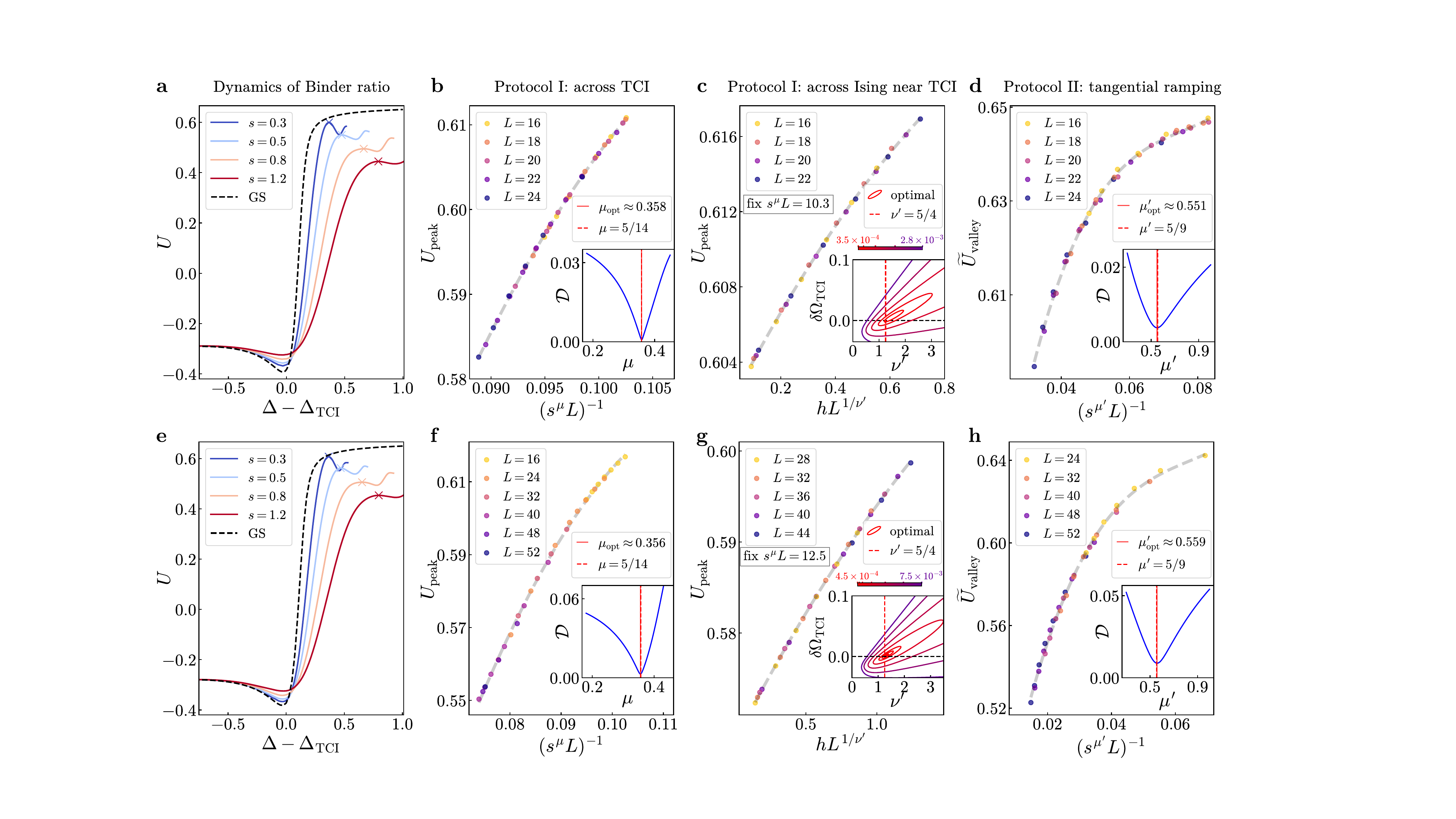}
    \caption{\textbf{KZ scaling of Binder ratio in the TCI case.} TEBD simulations using Hamiltonian \mbox{\eqref{eq:Hamiltonian}} with $N=2L$ atoms and $1/r^6$ interactions for (\textbf{a}--\textbf{d}). \textbf{a} Binder ratio $U$ as function of $g=\Delta-\Delta_\mathrm{TCI} \propto t$ for various ramping speed $s$ with $L=16$. The first peak is marked with a cross. The ground state Binder ratio for open boundary conditions (black dashed line) is included for reference. \textbf{b} KZ ramping across the TCI point, with data collapse achieved for $s^{\mu} L$. \textbf{c} KZ ramping near the TCI point, with data collapse achieved for $hL^{1/\nu'}$. \textbf{d} For tangential KZ, we achieve data collapse with $s^{\mu'} L$. In the insets, the data collapse figure of merit $\mathcal{D}$ is plotted against the corresponding critical exponents. \textbf{e}--\textbf{h} show analogous results using an effective spin-1 model with size $L$ for TEBD simulations. The gray dashed curves represent quadratic fits in (\textbf{b},\textbf{c},\textbf{f},\textbf{g}) and cubic fits in (\textbf{d},\textbf{h}).}
    \label{Fig:slow}
        \end{center}
\end{figure*}

Figure \ref{Fig:slow}a shows how the Binder ratio changes over time when ramping across the critical point at different speeds $s$. We observe that the Binder ratio oscillates rather than directly reaching a plateau, so we need to be cautious when using Binder data to determine scaling laws. The Binder curves at different speeds should show similar behavior due to the scaling invariance of the criticality. We choose the first peak after crossing the critical point ($t>0$) as a reference point for finite-size scaling. This is known as finite-time scaling \cite{gong2010finite,Huang2014FTS,shu2025emergent}, meaning that we do not need to wait for a long time for the quantity to saturate; instead, we use data that is relatively close to the critical point. This is crucial for tricritical physics, as data far from the critical point may not obey tricriticality, as we will see later.

We now discuss our first ramping scheme, Protocol I, which involves ramping across or near the tricritical point, as shown in Fig.~\ref{Fig:setup}c. Experimentally, this corresponds to ramping the detuning $\Delta$ within a range of fixed Rabi frequency $\Omega$ around $\Omega_\mathrm{TCI}$, with the deviation from the tricritical point parametrized by $h = \Omega - \Omega_\mathrm{TCI}$. While the tricritical point $\Omega_\mathrm{TCI}$ can be determined with high precision numerically, achieving this can be tricky experimentally. We will see that without knowing $\Omega_\mathrm{TCI}$ a priori, using scaling techniques we can still extract the critical exponents. A side product of our protocol is a method of finding $\Omega_\mathrm{TCI}$.

From a theoretical perspective, scanning near the tricriticality provides information about two relevant perturbations, resulting in both leading and subleading critical exponents \cite{Wilding1996,Kwak2015,Mathey2020,Ladewig2020}. In particular, the leading critical exponent, $\mu$, can be extracted by ramping the parameter $g$ at speed $s$, following the standard KZ scaling. 
Simultaneously, the parameter $h$ is associated with a subleading critical exponent, $\nu'$, with $[h]=1/\nu'$ (see Methods and Fig.~\ref{fig:central_charge} for more details). Note that $h^{\nu'} L$ forms a dimensionless combination. These critical exponents can be obtained from conformal field theory (CFT), summarized in Table~\ref{tab:cft}.

\begin{table}
\begingroup
\renewcommand{\arraystretch}{1.1}
\begin{ruledtabular}
\begin{tabular}{lccccc}
\textrm{Universality class} & $c$ & $\nu$ & $\mu$ & $\nu'$ & $\mu'$ \\
\colrule
two-leg Ising      & $1/2$  & 1      & $1/2$  & /    & / \\
two-leg TCI        & $7/10$ & $5/9$  & $5/14$ & $5/4$ & $5/9$ \\
three-leg Potts    & $4/5$  & $5/6$  & $5/11$ & /    & / \\
three-leg TCP      & $6/7$  & $7/12$ & $7/19$ & $7/4$ & $7/11$ \\
\end{tabular}
\end{ruledtabular}
\endgroup
\caption{The central charge and critical exponents of the universality classes.
The dynamical exponent $z$ for all cases is $1$, due to the emergent Lorentz symmetry.}
\label{tab:cft}
\end{table}

With these considerations in mind, we propose a dynamical finite-size scaling for Binder peaks near the tricritical point, described by the following equation
\begin{equation}
U_\mathrm{peak} = f(s^\mu L, h^{\nu'} L), \label{eq:scaling}
\end{equation}
where $f$ is a universal function.

We perform extensive numerical simulations of the dynamical evolution using the time-evolving block decimation (TEBD) algorithm with open boundary conditions~\cite{TEBD,Verstraete2004TEBD,keesling2019quantum,Chepiga2024KZ,zhang2025near}. Using the Binder peaks extracted from Fig.~\ref{Fig:slow}a for different system sizes $L$, ramping speeds $s$, and ramping paths parameterized by $h$, we verify Eq.~(\ref{eq:scaling}). In particular, we expect that for exponents $\mu$ and $\nu'$ that match the universal ones, $U_\mathrm{peak}$ as a function of $s^\mu L$ and $h^{\nu'} L$ will exhibit an optimal data collapse. Our numerical results lend full support for this expectation, as shown in the main figures of Fig.~\ref{Fig:slow}b, c. Figure~\ref{Fig:slow}b corresponds to $h = 0$, i.e., ramping across the tricritical point, and we fix $s^\mu L$ and collect data for a window of $\Omega$ in Fig.~\ref{Fig:slow}{c}. In the inset of Fig.~\ref{Fig:slow}{b}, we test different values of $\mu$ by a figure of merit $\mathcal{D} = \sqrt{\sum_i d_i^2}$, where $d_i$ is the vertical distance between the data point at $i$ and the smooth fitting curve value at the same horizontal value of $s^\mu L$. We plot $\mathcal{D}$ as a function of $\mu$, with the minimum corresponding to the optimal data collapse, which agrees closely with the CFT prediction. Next, in the inset of Fig.~\ref{Fig:slow}{c}, we show that we can obtain $\nu'$ even without the knowledge of $\Omega_\mathrm{TCI}$ by taking the latter as another optimization parameter. The contour of the figure of merit shows a good agreement with both $\nu'$ and the earlier used $\Omega_\mathrm{TCI}$, and $\delta\Omega_\mathrm{TCI}$ in the inset of Fig.~\mbox{\ref{Fig:slow}}{c} represents the difference from the value obtained from DMRG.

Additionally, we perform larger-scale simulations by making certain approximations. Since $r_2$ is sufficiently small, two atoms in a rung cannot be simultaneously excited to the Rydberg state. Hence, each rung can be recognized as a spin-1 mode (see the Methods section for more details).
With such a spin-1 representation, we can simulate much larger system sizes, more than double the original sizes, as shown in Fig.~\ref{Fig:slow}{f--h}. The conclusions drawn from the full $N=2L$ atom simulations remain unchanged, achieving the same universal scaling behavior.

From the numerical results, we conclude that this protocol provides a highly precise method for probing the leading and subleading critical exponents without the need to prepare critical ground states, bringing the experimental observation of the elusive tricritical point much closer to reality.\\ \\

\begin{figure*}[t!]
    \begin{center}  
    \includegraphics[width=\textwidth]{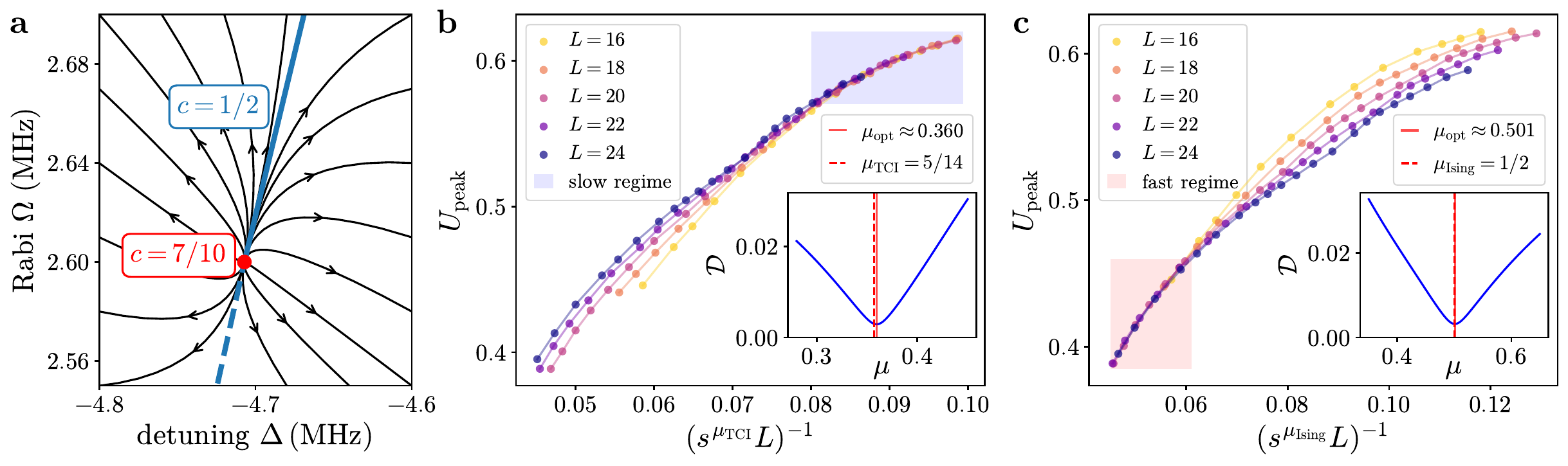}
    \caption{\textbf{RG perspective of KZ in the TCI case.} \textbf{a} RG flow (black lines) near the TCI. The blue line indicates the numerical phase boundary determined by Binder ratio and central charge (see Methods). \textbf{b, c} Ramping with a broad range of ramping speed $s$. Both figures are plotted with the same data but with different scalings of $s$. For slower ramping (upper right region in (\textbf{b})), the TCI critical exponent allow the best data collapse. For faster ramping (lower left region in (\textbf{c})), the Ising critical exponent allow the best data collapse.} 
    \label{Fig:fast}
        \end{center}
\end{figure*}

\noindent \textbf{Tangential KZ scaling} \\
\noindent The two-dimensional phase diagram with a curved phase boundary allows a novel type of KZ scaling, which we dub ``tangential'' to highlight how the dynamical path touches the phase boundary. The conceptual novelty here lies in the fact that throughout the ramping one stays exclusively in the disordered phase and never enters the ordered phase, unlike previous KZ protocols. This fact leaves uncertain some heuristic arguments commonly found in the literature, where domain walls or generally topological defects that are only well defined in the symmetry-breaking phase play a key role. Nevertheless, scaling relations that do not depend on detailed modeling of the dynamics remain valid. 

In the tangential ramping, we are effectively ramping the subleading operator, with the leading operator vanishing (see Methods for more details). Hence, the scaling relation is modified to
\begin{equation}
U=f(s^{\mu'} L) \text{, with } \mu' = (z + 1/\nu')^{-1},\label{eq:scaling2}
\end{equation}
which provides a direct probe of the subleading exponent. It is known previously that the Binder ratio works the best in the ordered phase but scales not as well in the disordered phase \cite{king2023quantum}. To remedy this, we introduce a disordered Binder ratio $\widetilde{U}=1- \frac{1}{3}\langle \widetilde{M}^4\rangle/\langle \widetilde{M}^2\rangle^2$. Here the disorder parameter is defined as
\begin{equation}
\widetilde{M}=\sum_i\prod_{j\geq i}\operatorname{SWAP}_j,
\end{equation}
where $\operatorname{SWAP}$ interchanges the atoms of each rung. The operator can be equivalently written as $\widetilde{M}=\operatorname{SWAP}_1(I_2+\operatorname{SWAP}_2(\cdots
+\operatorname{SWAP}_{L-1}(I_L+\operatorname{SWAP}_L)\cdots))$, which facilitates an efficient matrix-product operator (MPO) calculation. Unlike the ordered Binder ratio, which we choose the peak, the disordered Binder ratio exhibits a valley when tangentially passing through the tricritical point, as shown in Supplementary Note 3. The disordered Binder valleys are collected, and the data collapse is shown in Fig.~\ref{Fig:slow}{d}, which is fully consistent with the scaling relation in Eq.~\eqref{eq:scaling2}. \\

\noindent \textbf{Dynamical Zamolodchikov's $c$-theorem} \\
\noindent Quantum dynamics can be roughly classified by its characteristic time scale, with adiabaticity and quench the two opposite limits. In the case of KZ, a natural parameter is the ramping speed $s$, and $s$ is generally assumed to be small so that the physics is well controlled by the critical point. In previous discussions, this is also the regime we focused on. Now we go slightly beyond and focus on an intermediate-speed ramping. 

Heuristically, upon inspecting the phase diagram in Fig.~\ref{Fig:setup}{c} with a 0d tricritical point and a 1d critical line, we expect only the neighboring region of the tricritical point to be  controlled by its universal property, and as one moves far away the Ising/Potts criticality takes over. The argument can be made more precise by examining the RG flow around the tricritical point, see Fig.~\ref{Fig:fast}{a}. One sees that far away from the tricritical point, the RG flow generically takes a detour near the critical line, thus inheriting criticality from the latter~\cite{wang2025kink,li2025random}. 

Returning to dynamics, it follows that the tricritical exponents can only be observed with sufficiently slow ramping. This is evident in Fig.~\ref{Fig:fast}{b}, where the upper-right corner of the figure shows a good data collapse when the horizontal axis is scaled by the tricritical exponent $\mu_\text{TCI}$. For faster ramping speeds, data for different system sizes spread out. However, if we rescale the horizontal axis by the critical exponent $\mu_\text{Ising}$, as shown in Fig.~\ref{Fig:fast}{c}, the data collapse is restored in the fast ramping regime. This indicates that in this regime, the emergent tricritical Ising is not observed, but the original Ising criticality manifests instead. This result holds generally for different ramping directions through the tricritical point. Specifically, by choosing a direction in which the tricriticality extends the furthest in the ground state (Supplementary Note 4), one might expect that only TCI would be observed in this regime. Nevertheless, we still observe Ising criticality dominating the dynamics. The appearance of two distinct critical exponents across different ramping speed ranges is a generic feature that reflects this rich renormalization group structure, as we also demonstrate for the three-leg case (see Supplementary Note 2).

These results are reminiscent of the celebrated Zamolodchikov's $c$-theorem. The theorem states that along an RG flow the central charge $c$ has to decrease. This physics is again well represented by the case of tricritical points in the ground state, where two different criticalities come into play, and is best illustrated by the RG flow in Fig.~\ref{Fig:fast}{a}. Although the KZ protocols do not directly measure the central charge, our observed crossover behavior still largely conforms to the same phenomenology, namely a flow from tricriticality (high $c$) to the Ising/Potts criticality (low $c$).

\noindent {\large \textbf{Discussion}}\\
\noindent In this work, we systematically investigate the KZ scaling near tricritical points, using two- and three-leg Rydberg ladders. Through both conventional and tangential ramping protocols near the tricritical point, we accurately determine the critical exponents associated with the two relevant operators, in close agreement with their CFT predictions at the exact tricriticality. Additionally, we uncover a dynamical analog of Zamolodchikov's $c$-theorem in an intermediate-speed ramping regime.

We anticipate an imminent experimental realization of our proposals, particularly for the two-leg model which only requires a planar geometry and harbors TCI criticality. Looking forward, the TCI is closely related to spacetime supersymmetry---a long-sought concept in theoretical physics. In our setup, the critical exponents $\nu$ and $\nu'$ correspond to the bosonic sector of the two lowest weight symmetry-preserving fields, $\epsilon$ and $\epsilon'$. These fields exhibit power-law correlations, $\langle \epsilon_i \epsilon_j \rangle \sim |i-j|^{-2\Delta_\epsilon}$, with $\Delta_\epsilon = 2 - 1/\nu$ (and similarly for $\epsilon'$, giving $\Delta_{\epsilon'}$).  Notably, $\Delta_{\epsilon'} - \Delta_{\epsilon}= 1/\nu-1/\nu' = 1$. This relationship is no coincidence: the fermionic mode $\psi$ is linked to these fields through the supersymmetry generator, resulting in the relation $\Delta_\psi = \Delta_{\epsilon} + 1/2 = \Delta_{\epsilon'} - 1/2$, illustrating the structure of fermion bridging the two bosons. Hence, our protocol offers an effective method of corroborating this nontrivial ramification of emergent supersymmetry. On the other hand, directly measuring the corresponding fermion $\psi$ dynamically remains an open question, as it would likely require ramping fermionic operators \cite{Bravyi2002,li2024supersymmetry}. This introduces an interesting direction for future research, which we intend to explore further.

Note added: In a recent post \cite{yin2025driven}, the authors also investigate the KZ mechanism at a supersymmetric tricritical point with two relevant directions. Although their model differs from ours, the same universality emerges, leading to overlapping and mutually corroborating findings.\\

\noindent {\large \textbf{Methods}}\\
\noindent \textbf{Ground state phase diagrams} \\
\noindent We describe the method used to obtain the ground state phase diagrams for the two- and three-leg ladders. In both cases, we employ DMRG to compute the ground state $|\psi\rangle$.

For each Rabi frequency $\Omega$, we sweep the detuning $\Delta$ to calculate the Binder ratio of the ground state under periodic boundary conditions (PBC), $U_\text{GS}(L,\Omega,\Delta) =1- \frac{1}{3}\langle \psi | M^4 |\psi\rangle/\langle \psi| M^2 |\psi \rangle^2$. For different system sizes $L$, the Binder curve $U_\text{GS}$ versus $\Delta$ crosses at a single point, which is used to determine the transition point. By collecting the transition detuning values along with their corresponding Rabi frequencies, we obtain the phase boundary.

Next, we identify the tricritical point by calculating the central charge $c$ along the phase boundary. The maximum value of $c$ identifies the tricritical point \cite{li2024supersymmetry}. Specifically, we calculate the entanglement entropy along the phase boundary. The entanglement entropy $S(l)$ of a subsystem with length $l$ is defined as $S(l) = -\operatorname{Tr} \rho(l) \log \rho(l)$, where $\rho(l) = \operatorname{Tr}_{L \setminus l} |\psi\rangle \langle \psi|$ is the reduced density matrix obtained from the ground state $|\psi\rangle$ calculated under PBC. The central charge of the state is then obtained by fitting the entanglement entropy to \cite{Vidal2003,Calabrese-Cardy_2004}
\begin{equation}
S(l) = \frac{c}{3} \log \left[\sin\left(\frac{\pi l}{L}\right)\right] + \text{const}.
\end{equation}

\begin{figure}[t!]
    \begin{center} 
    \includegraphics[width=1\columnwidth]{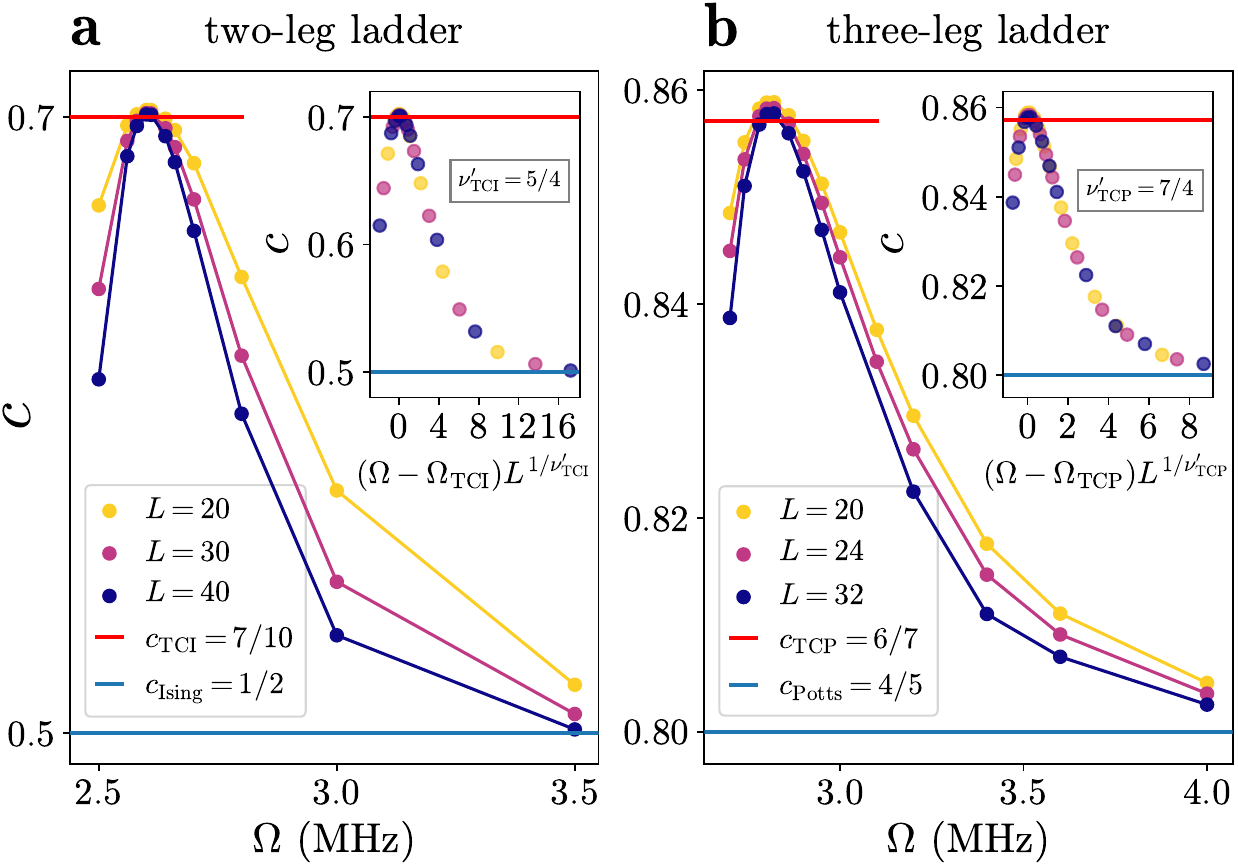}
    \caption{Central charge $c$ vs Rabi frequency $\Omega$ for two-leg (\textbf{a}) and three-leg (\textbf{b}) cases, using cesium 70S and sodium 69S. Insets show data collapse with horizontal axis $hL^{1/\nu'}$, validating the scaling proposal of Eq.~\eqref{eq:scaling}.} \label{fig:central_charge}
    \end{center}
\end{figure}

For the two-leg model, we use cesium 70S (a-type) and sodium 69S (b-type) as the two species of atoms in the main text. The interactions between atoms are van der Waals $C/r^6$, with interaction constants $(C_\mathrm{aa}, C_\mathrm{bb}, C_\mathrm{ab}) = (650, 1570, -614) \, \unit{GHz.\micro\meter^6}$. The Rydberg blockade is manifest with $r_2 = 5 \unit{\micro\meter}$. Then $r_1 = 7 \unit{\micro\meter}$ sets the attractive energy scale to $5.2 \unit{MHz}$. This energy scale controls the position of the tricritical point $\Omega_\text{TCI}$ and $\Delta_\text{TCI}$, allowing one to tune it to a desired value in experiments by adjusting $r_1$. With these parameters, we plot how $c$ changes along the phase boundary characterized by $\Omega$, near the tricritical regime, as shown in Fig.~\ref{fig:central_charge}{a}. The central charge peaks at $\Omega_\text{TCI}\approx 2.6 \,\text{MHz}$ with $c=7/10$, and gradually decreases to $1/2$ in the direction toward the standard second-order line. The tangential scaling $\nu'$ is explicitly demonstrated in the inset, through proper data collapse with the horizontal axis $hL^{1/\nu'}$. The numerical phase diagrams are presented in Supplementary Note 1 and share the same features as Fig.~\mbox{\ref{Fig:setup}}{c} in the main text.

Besides the full-range van der Waals calculation mentioned above, we also performed finite-range interaction calculations (i.e., truncating longer-range interactions). In the latter case, we include interactions up to: ${}^{\Circle}_{\Circle}{}^{\Circle}_{\Circle}{}^{\CIRCLE}_{\Circle}{}^{\Circle}_{\Circle}{}^{\Circle}_{\Circle}$, where $\CIRCLE$ represents the reference atom and $\Circle$ represents atoms whose interactions with the reference atom are included. We find no differences between the full-range and such finite-range DMRG results. The reason for checking finite-range interactions is that in dynamical simulations, including all-range interactions makes Trotter decomposition very challenging. Hence, in the dynamics calculations, as detailed in the following section, we include only this finite range of interactions.

For the three-leg ladder case, we also performed the same calculation with the same parameters of cesium 70S and sodium 69S as used in the two-leg case, by simply adding one more atom to each rung, as shown in Fig.~\ref{Fig:setup}{b}. The central charge result is similar to its two-leg cousin, except that the numerical values of the central charge and critical exponents follow those of Potts and tricritical Potts, see Fig.~\ref{fig:central_charge}{b}. However, it is quite heavy to perform dynamical evolution with the full setup. Hence, to demonstrate the universality feature of the physics, we also propose an effective model, as detailed in the following section. The same universality is found, with the phase diagram presented in Supplementary Note 1.\\

\noindent \textbf{The effective model for the two-leg ladder} \\
\noindent Here, we present an effective description of the two-leg ladder when the blockade condition is well satisfied. In each rung, two spin-1/2 atoms form a Hilbert space of four states. If both atoms cannot be simultaneously excited due to the blockade, then only three states remain accessible, and each rung can be effectively described as a spin-1 mode. To be more specific, we label the atoms on the upper rung as $i=1,3,5,\ldots,2L-1$ and the lower rung atoms as $j=2,4,6,\ldots,2L$. Each rung is labeled by $I=1,2,3,\ldots,L$, which also labels the corresponding spin-1. In this representation, the Hamiltonian~\mbox{\eqref{eq:Hamiltonian}} can be written in terms of $S^x$ and $S^z$ operators, with the following correspondence:
\begin{equation}
\begin{aligned}
\Omega(\sigma^x_i + \sigma^x_j) &\longleftrightarrow \sqrt{2}\Omega S^x_I\\
\Delta(n_i + n_j) &\longleftrightarrow \Delta S^z_I\\
V_1^{\mathrm{ab}}(n_in_{i+1}+n_{j}n_{j+1}) &\longleftrightarrow \frac{V_1^{\mathrm{ab}}-V_2^{\mathrm{ab}}}{2}S^z_IS^z_{I+1}\\
V_2^{\mathrm{ab}}(n_in_{j+1}+n_{i+1}n_{j}) &\longleftrightarrow \frac{V_1^{\mathrm{ab}}+V_2^{\mathrm{ab}}}{2}(S^z_I)^2 (S^z_{I+1})^2\label{eq:spin-1}
\end{aligned}
\end{equation}
where the attractive interaction between dual-species atoms is given by $V_1^{\mathrm{ab}}=C_{\mathrm{ab}}/r_1^6$ (upper-upper/lower-lower interaction) and $V_2^{\mathrm{ab}}=C_{\mathrm{ab}}/(r_1^2+r_2^2)^3$ (upper-lower/lower-upper interaction). Longer-range interactions between $S_I^z$ and $S_{I+2}^z$ can also be included, with repulsive interaction strengths $V_1^{\mathrm{aa}}=C_{\mathrm{aa}}/(2r_1)^6$ and $V_2^{\mathrm{aa}}=C_{\mathrm{aa}}/((2r_1)^2+r_2^2)^3$, and similar expressions for $V_1^{\mathrm{bb}}$ and $V_2^{\mathrm{bb}}$.

The same universality is observed in the ground state phase diagram, with the non-universal feature that the position of the tricritical point is shifted by approximately $5\%$ when only attractive effects are included. When longer-range repulsive interactions are also included, the position of the tricritical point shifts by approximately $0.6\%$ compared to the full range spin-$1/2$ model.\\

\noindent \textbf{The effective model for the three-leg ladder} \\
\noindent The effective Hamiltonian for the three-leg ladder reads
\begin{equation}
H=\sum_i\frac{\Omega}{2}X_i-\Delta n_i -Jn_in_{i+1}.\label{eq:3leg}
\end{equation}
In this effective model, by enforcing Rydberg blockade, each site has four states $|0\rangle$, $|a\rangle$, $|b\rangle$ and $|c\rangle$, corresponding to all atoms in the ground state or one of the atoms on a rung excited to the Rydberg state, respectively. $X=\sum_{\alpha=a,b,c}|\alpha\rangle\langle0|+\mathrm{h.c.}$ couples the ground state and the Rydberg state as usual, and $n=\sum_{\alpha=a,b,c}|\alpha\rangle\langle\alpha|$ projects to the Rydberg states. Inter-rung attractive interactions are encompassed in a coupling constant $J$, which we set to 1 as the energy scale.

To properly work with the three-leg ladder, we need to redefine several physical quantities. The order parameter is defined by
\begin{equation}
M_3=\sum_i n_{ia}+e^{2\pi\ri/3}n_{ib}+e^{4\pi\ri/3}n_{ic},
\end{equation}
and the disorder operator reads
\begin{equation}
\widetilde{M}_3=\sum_i\prod_{j\geq i}\operatorname{CYCL}_j,
\end{equation}
where $\operatorname{CYCL}=|0\rangle\langle 0| + |b\rangle\langle a|+|c\rangle\langle b|+|a\rangle\langle c|$ permutes the atoms along the rungs. A numerics-friendly version is $\widetilde{M}_3=\operatorname{CYCL}_1(I_2+\operatorname{CYCL}_2(\cdots
+\operatorname{CYCL}_{L-1}(I_L+\operatorname{CYCL}_L)\cdots))$. The Binder ratio can then be defined in the same way as described in the main text.\\

\noindent \textbf{Numerical simulation of KZ dynamics} \\
\noindent We employ the TEBD algorithm to simulate the KZ dynamics. We include finite-range interactions and use Trotter decomposition for the unitary evolution with open boundary conditions. During the ramping process, the time step is fixed at $\Delta t$. To ensure the convergence of our results, we test different values of $\Delta t = 0.0004, 0.0002, 0.0001 \unit{\micro s}$, always obtaining consistent results. All ramping speeds in the main text are given in units of $\unit{MHz/\micro s}$. The singular value decomposition truncation cutoff of $5 \times 10^{-12}$ is used. With this approach, the time-dependent wavefunction $|\psi(t)\rangle$ is obtained, and various physical operators can be measured from this wavefunction to extract the quantities discussed in the main text.

For the two-leg case, as the DMRG results suggest, finite-range truncation of the $C/r^6$ interactions in Hamiltonian \mbox{\eqref{eq:Hamiltonian}} still exhibits proper tricriticality. In the dynamical simulation, when implementing the Trotter decomposition, interactions are included for every group of six atoms. For example, the $\CIRCLE$ atom participates in three different groupings:  ${}^{\Circle}_{\Circle}{}^{\Circle}_{\Circle}{}^{\CIRCLE}_{\Circle}$, ${}^{\Circle}_{\Circle}{}^{\CIRCLE}_{\Circle}{}^{\Circle}_{\Circle}$, ${}^{\CIRCLE}_{\Circle}{}^{\Circle}_{\Circle}{}^{\Circle}_{\Circle}$. Boundary atoms are handled appropriately under open boundary conditions by adjusting the grouping scheme near the edges, following approaches similar to Refs.~\mbox{\cite{keesling2019quantum,zhang2025near}}. These simulations produce the results shown in Fig.~\mbox{\ref{Fig:slow}}{a}--{d}. Additionally, we employ the spin-1 representation [cf. Eqs.~\mbox{\eqref{eq:spin-1}}] for TEBD simulations. This approach reduces the dimension of the local Hilbert space, making dynamical simulations feasible for much larger system sizes, as demonstrated in Fig.~\mbox{\ref{Fig:slow}}{e}--{h}. Importantly, the same universal dynamical scalings are uncovered in both methods.

For the three-leg case, we use Hamiltonian \mbox{\eqref{eq:3leg}} for TEBD simulations, and results are summarized in Supplementary Note 2.
\\

\noindent \textbf{Critical exponents and scaling dimensions} \\
\noindent In the main text, we identified two critical exponents $\nu$ and $\nu'$ near the tricriticality. In the CFT language, these exponents correspond to two symmetry-preserving relevant conformal fields, $\epsilon$ and $\epsilon'$, in tricritical Ising/Potts CFT, with scaling dimensions $\Delta_\epsilon$ and $\Delta_{\epsilon'}$, respectively.

The $\epsilon$ field is very similar to its counterpart at the parent critical line. It drives the system between disordered and ordered phases. For a shifted parameter $g$, the field couples to it with an additional action term compared to the critical point value: $S = \int \mathrm{d}x \, \mathrm{d}\tau \, g \cdot \epsilon$. Near criticality, scaling invariance is still preserved and action is unchanged under the following scaling transformations with ratio $b$:
\begin{equation}
x \to b^{-1}x,\,\, \tau \to b^{-z}\tau,\,\, g \to b^{1/\nu}g,\,\, \epsilon \to b^{\Delta_\epsilon}\epsilon
\end{equation}
This immediately gives the relation between the critical exponents that can be measured in experiments and the scaling dimension of the coupled conformal field: $\Delta_\epsilon + 1/\nu = 1 + z$ [cf. Table~\ref{tab:cft} and Table~\ref{tab:cft2}].

\begin{table}[t]
\begingroup
\renewcommand{\arraystretch}{1.2}
\begin{ruledtabular}
\begin{tabular}{ccccccc}
\textrm{Tricritical} & $\Delta_\epsilon$ & $\Delta_{\epsilon'}$ & $a_1$ & $a_2$ & $b_1$ & $b_2$ \\
\colrule
TCI & $1/5$ & $6/5$  & 0.252 & $-0.299$ & 0.347 & 0.073 \\
TCP & $2/7$ & $10/7$ & 0.465 & $-0.174$ & 0.281 & 0.053 \\
\end{tabular}
\end{ruledtabular}
\endgroup
\caption{Conformal dimensions and expansion coefficients of TCI and TCP.}
\label{tab:cft2}
\end{table}

\begin{figure}
    \begin{center}  
    \includegraphics[width=0.85\columnwidth]{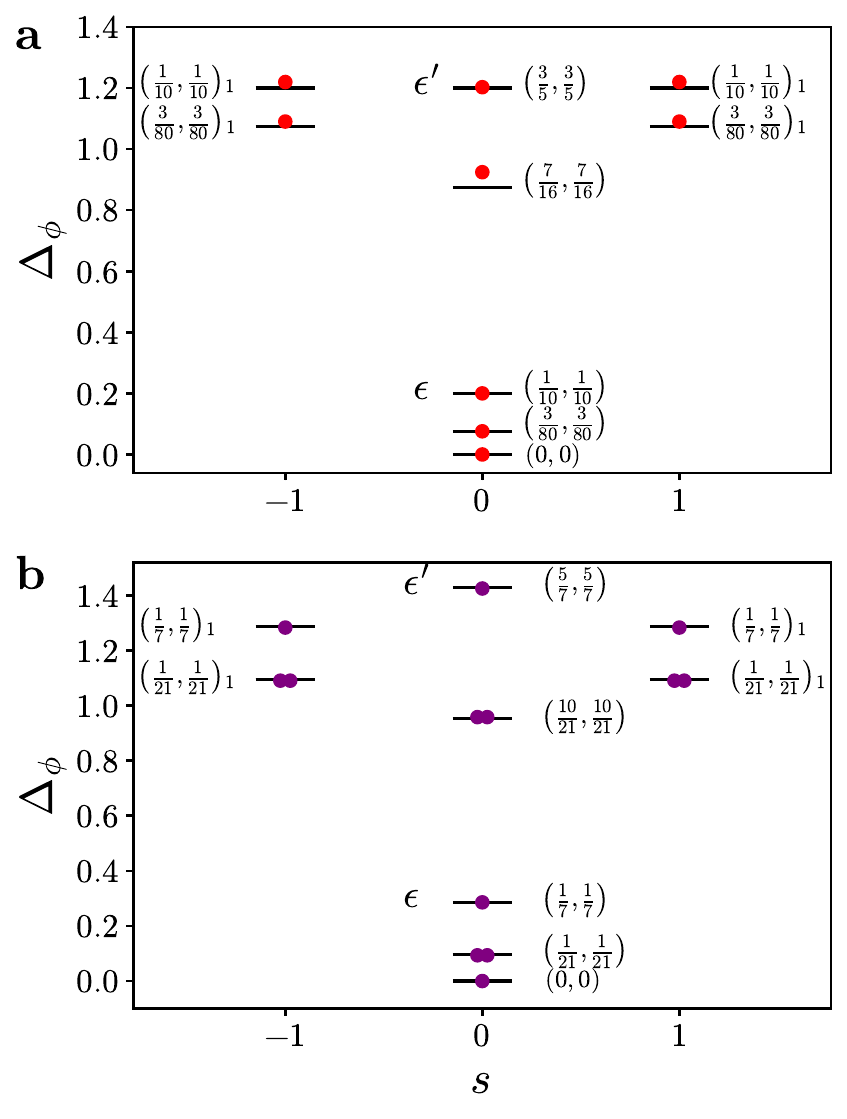}
    \caption{The low-energy spectrum of TCI (\textbf{a}) and TCP (\textbf{b}). The states are identified by their conformal weights $(h,\bar{h})$, with conformal dimension (microscopically, energy) $\Delta_\phi=h+\bar{h}$ and conformal spin (microscopically, momentum) $s=h-\bar{h}$.} 
    \label{fig:state}
        \end{center}
\end{figure}

The $\epsilon$ field, as the most relevant operator, is responsible for the spontaneous symmetry breaking. However, in the phase diagram spanned by two parameters, there is a direction that does not lead to either phase---namely, the direction parallel to the local phase boundary. In this direction, the perturbation involving $\epsilon$ is completely canceled, and the subleading field $\epsilon'$ takes charge. We denote the angle of the phase boundary relative to the detuning direction as $\theta$, and the deviation in this direction from the tricritical point as $h_\parallel$. This angle $\theta$ can be determined by expanding lattice operators in terms of conformal fields, as illustrated in the next subsection. $h_\parallel$ refers to the strength of deviation from the tricritical point, with the equilibrium critical exponent defined as $\nu'$. Since $h_\parallel$ couples to $\epsilon'$, which contributes most significantly to the action, following the same argument gives $\Delta_{\epsilon'} + 1/\nu' = 1 + z$. In the main text, we define $h=\Omega-\Omega_\mathrm{TCI}$ to control the deviation from the tricritical point, which has the relation $h=h_\parallel\sin(\theta)$. We use the dimension $[h]=1/\nu'$ in this sense, as verified by the central charge scaling along the phase boundary shown in the insets of Fig.~\ref{fig:central_charge}.

With these equilibrium features near the tricritical point, we therefore propose a ``tangential ramping'' protocol that tangentially approaches the tricritical point. In this direction, before reaching the near-tricritical regime, the energy gap remains finite. Near the tricritical point, the KZ mechanism becomes relevant due to the gapless nature, with the scaling entirely controlled by $\epsilon'$ and thus enabling measurement of $\mu'$.\\

\noindent \textbf{Expansion of lattice operators by conformal fields} \\
\noindent In this section, we explain how we obtain the renormalization group flow at the tricritical point shown in Fig.~\ref{Fig:fast}{a}. We relate lattice operators $\sigma^x$, $n$ [cf. Eq.~\mbox{\eqref{eq:Hamiltonian}}] to conformal fields. From symmetry considerations, we have
\begin{equation}
\begin{aligned}
\sigma^x&=\langle \sigma^x\rangle+ a_1\epsilon +a_2\epsilon'+\dots\\ 
n\,\,&=\,\,\langle n\rangle\,+ b_1\epsilon \,+b_2\epsilon'+\dots\label{eq:expan}
\end{aligned}
\end{equation}
We note that $\sigma^x$ and $n$ are coupled to $\Omega$ and $\Delta$, respectively. Once we determine these coefficients, and hence the relative strength between $\epsilon$ and $\epsilon'$, we can establish (i) the direction of the phase boundary (where the $\epsilon$ component vanishes) and (ii) the generic RG flow trajectory in the phase diagram with parameters $\Omega$ and $\Delta$. We explain how to extract these coefficients numerically below.

The main analytical tool we use is the state--operator correspondence in CFT. Roughly speaking, this relates the energy eigenstates of a microscopic model to the primary fields of the field theory. Two important consequences follow. First, the energies of the lowest eigenstates are given by $E_\phi=\Delta_\phi+n$,
where $\Delta_\phi$ is the conformal dimension of the field $\phi$ and $n$ labels the descendant level. Here we have shifted and rescaled the energy spectrum so that the ground state has zero energy and descendant states have unit energy intervals. We can thus identify the particular states $|\epsilon\rangle,|\epsilon'\rangle$, see Fig.~\ref{fig:state}. Second, we have the matrix element $\langle\phi|\phi|0\rangle=(2\pi/L)^{\Delta_\phi}$.
Combined with the expansion, Eq.~\eqref{eq:expan}, we can extract the four coefficients of interest. The conformal dimensions and the numerical results are collected in {Table \ref{tab:cft2}}. This is the data we used to plot the RG flow in Fig.~\ref{Fig:fast}{a}. We also mention that the densest direction of the RG trajectory lines coincides with the DMRG phase boundary calculation (blue line, obtained using the Binder ratio), showing good agreement in determining the tangential direction.\\

\noindent {\bf Acknowledgements}\\
We are grateful to Hui Zhai for collaboration on related projects and stimulating ideas. We thank Wenjun Zhang for helpful discussions on the experimental details. We thank Shuai Yin for reading our manuscript and providing valuable feedback. This work is supported by National Natural Science Foundation of China under Grant No.~12504307 (C.L.). C.L. is also supported by Tsinghua University Dushi program. H.W. is supported by China Postdoctoral Science Foundation under Grant No.~2024M751609 and Postdoctoral Fellowship Program of CPSF under Grant No.~GZC20231364.
The DMRG and TEBD calculations are performed using the ITensor library \cite{ITensor}.\\

\noindent {\bf Author contributions}\\
H.W., X.L., and C.L.~contributed to developing the theory. H.W. performed the numerical simulations. All authors analyzed the data and wrote the manuscript.\\

\noindent {\bf Data availability}\\
The data generated in this study have been deposited in the Zenodo database under identifier \href{https://doi.org/10.5281/zenodo.17385416}{https://doi.org/10.5281/zenodo.17385416}.\\

\noindent {\bf Code availability}\\
The numerical codes used in this study are available from the corresponding author upon request.\\

\bibliography{TC_KZ.bib}

\begin{onecolumngrid}

\clearpage
\subsection*{\large Supplementary Information for \\ ``Tricritical Kibble--Zurek scaling in Rydberg atom ladders''}
\normalsize
\begin{center}
{Hanteng Wang, Xingyu Li, and Chengshu Li\texorpdfstring{\,\textsuperscript{\hyperlink{supp-email}{*}}}{}\\
\emph{Institute for Advanced Study, Tsinghua University, Beijing, 100084, China}}
\end{center}

\setcounter{equation}{0}
\setcounter{figure}{0}
\setcounter{table}{0}
\setcounter{section}{0}
\setcounter{page}{1}
\renewcommand{\tablename}{Supplementary Table}
\renewcommand{\figurename}{Supplementary Fig.}
\renewcommand{\thefigure}{S\arabic{figure}} 

\makeatletter
\section*{Supplementary Note 1: Ground state properties}
We present the numerical results for the ground state phase diagram of both the two-leg and three-leg models in Supplementary Fig.~\ref{fig:pd}. The phase boundary is determined by Binder ratio crossings, and the tricritical point is identified by the peak of the central charge, as shown in Fig.~\ref{fig:central_charge} in the main text.

\section*{Supplementary Note 2: More on numerical results for KZ dynamics}
We present the numerical results of all three KZ protocols for the three-leg ladder in Supplementary Fig.~\ref{Fig:Potts_disorder} and Supplementary Fig.~\ref{Fig:TCP_Potts}. All results show behavior nearly identical to those observed for the two-leg ladder, indicating that our proposals and findings are equally valid beyond the TCI case.

\section*{Supplementary Note 3: More on tangential ramping}
Using phase boundary data, we can determine the tangential line for the tricritical point, denoted as $\theta_t$ for the angle. For the TCI, $\theta_t \approx 70^\circ$, and for the TCP, $\theta_t \approx 50^\circ$. These results are consistent with CFT calculations at the tricritical point, as shown in Fig.~3a and Supplementary Fig.~\ref{Fig:TCP_Potts}a. With the tangential line determined, we prepare an initial state with zero Rabi frequency and then ramp both the Rabi frequency and detuning simultaneously. The coordinates of the ramping line on the phase diagram $(\Delta, \Omega)$ follow $\Delta = \Delta_\mathrm{TC} + st\cos\theta_t$, $\Omega = \Omega_\mathrm{TC} + st\sin\theta_t$, where $s$ is the ramping speed.

Using the disorder parameter, we plot how its corresponding Binder ratio $\widetilde{U}$ evolves over time as shown in Supplementary Fig.~\ref{Fig:Potts_disorder}a. It starts at $2/3$ and then decreases. After tangentially touching the tricritical point, the Binder ratio exhibits a valley structure, which we use as the reference point for the subsequent scaling procedure. Figure 2d and Supplementary Fig.~\ref{Fig:Potts_disorder}d show decent data collapse for this ramping protocol, allowing us to extract the subleading critical exponents $\mu'$.\\

\section*{Supplementary Note 4: Ground state scalings and the sweet-spot direction}
In this section, we discuss critical scalings near the tricritical point in the ground state. For a generic (i.e. not along the phase boundary), we expect $\xi_\mathrm{GS}\propto |\Delta-\Delta_c|^{-\nu}$. As one moves away from the tricritical point, the scaling generically does not hold. This can be seen from the finite-size scaling of the correlation length, expressed as $\xi_\mathrm{GS}/L = \widetilde{\mathcal{F}}((\Delta - \Delta_\mathrm{TCI})L^{1/\nu})$, where $\xi_\mathrm{GS}$ can be extracted from the \emph{disconnected} two-point correlator. However, by numerically exploring various directions emanating from the tricritical point, we find a sweet spot direction where the scaling behavior extends much farther than other directions, as shown in Supplementary Fig.~\ref{fig:scaling}. We emphasize that this direction does not originate from universality, but offers a window that enables the study of criticality away from the critical point. Indeed, it is made good use of when we explore intermediate-speed ramping as detailed in the main text. Interestingly, along this direction where the ground state is controlled by a single scaling law, KZ dynamics reveals a crossover effect: at slow speeds, tricritical scaling is observed, while at faster speeds, Ising/Potts scaling is recovered. This differs from cases where dynamical scaling is inherited from equilibrium scaling. Our observed crossover is a purely dynamical phenomenon arising from the interplay between two relevant operators.\\ \\ \\

\par\noindent\rule{0.1\textwidth}{0.5pt}\\

\noindent\hypertarget{supp-email}{}%
{\footnotesize \textsuperscript{*}\,\href{chengshu@mail.tsinghua.edu.cn}{chengshu@mail.tsinghua.edu.cn} \par}

\begin{figure*}[b!]
    \begin{center}  
    \includegraphics[width=0.7\columnwidth]{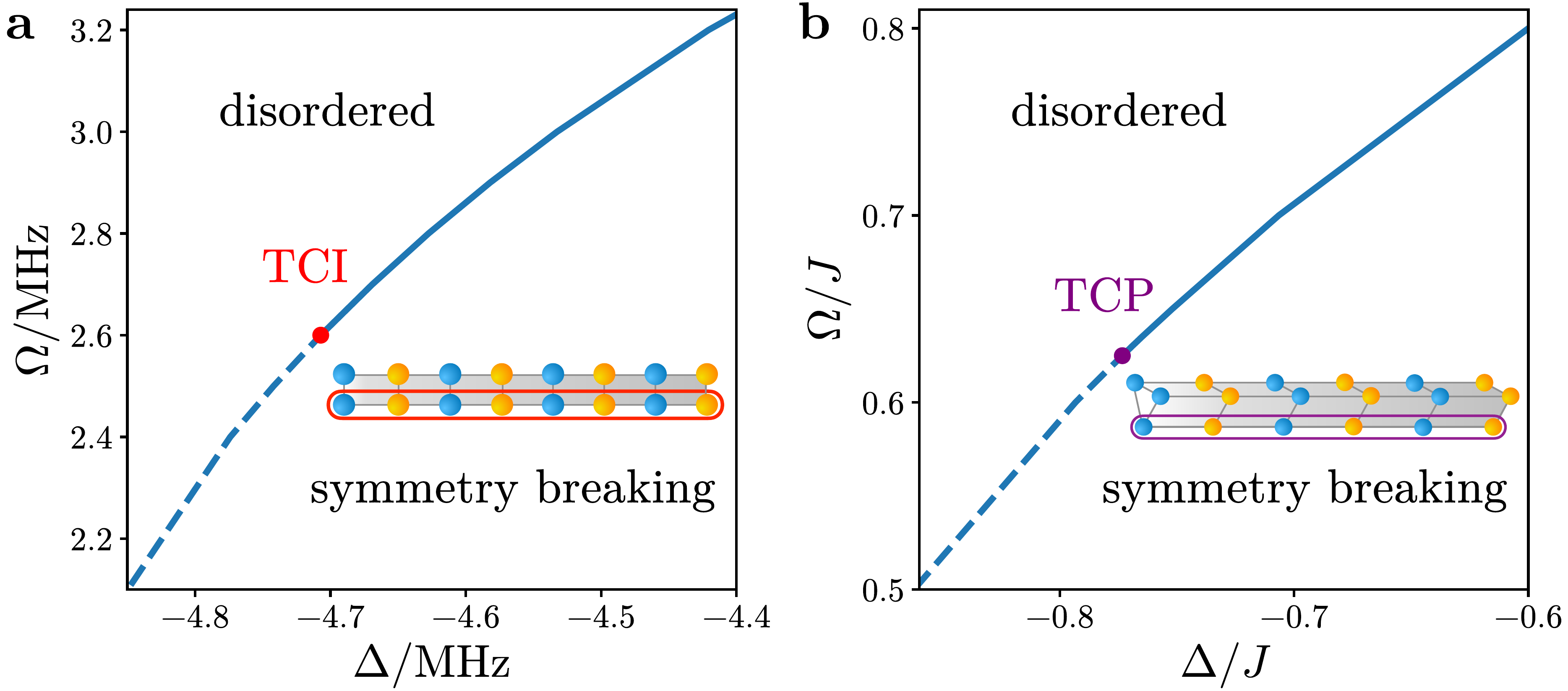}
    \caption{The phase diagram for two-leg (\textbf{a}) and three-leg (\textbf{b}) Hamiltonian, with tuning parameters $\Omega$ and $\Delta$.} 
    \label{fig:pd}
        \end{center}
\end{figure*}

\begin{figure*}[t!]
    \begin{center}  
    \includegraphics[width=1\textwidth]{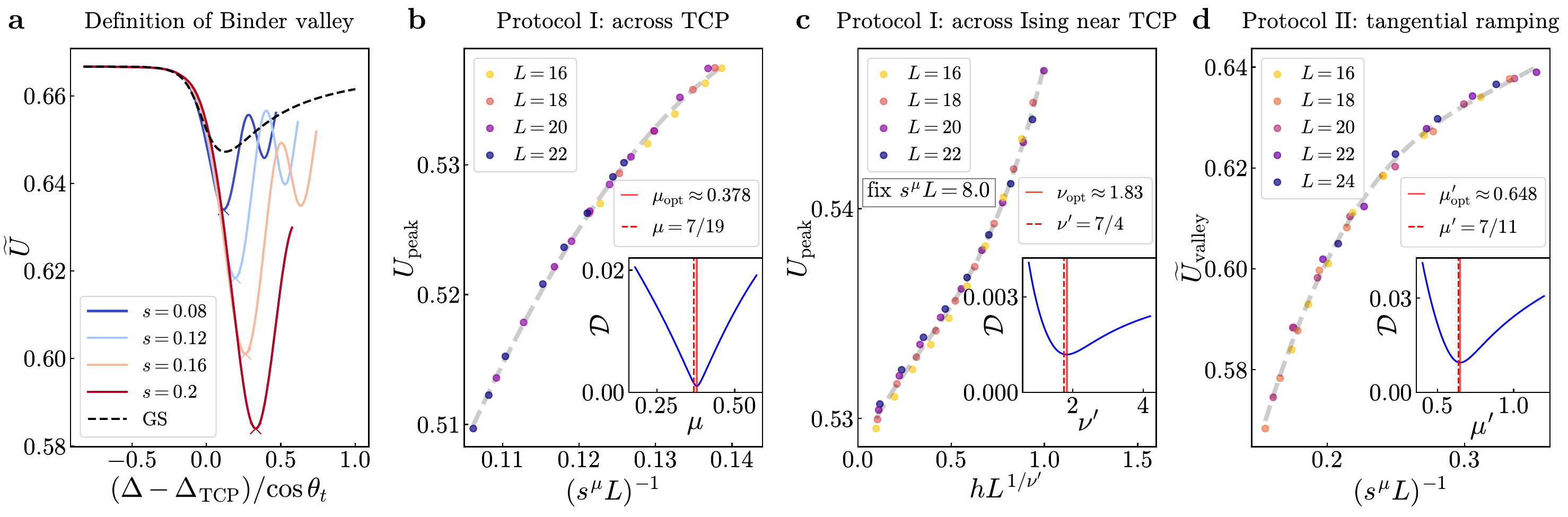}
    \caption{\textbf{a}, Binder ratio $\widetilde{U}$ vs $(\Delta-\Delta_\mathrm{TCP})/\cos\theta_t\propto t$ defined in tangential case. The first valley is marked with a cross. \textbf{b}, KZ ramping across the TCP point. \textbf{c}, KZ ramping near the TCP point. \textbf{d}, Tangential KZ for TCP.} 
    \label{Fig:Potts_disorder}
        \end{center}
\end{figure*}

\begin{figure*}[t!]
    \begin{center}  
    \includegraphics[width=1\textwidth]{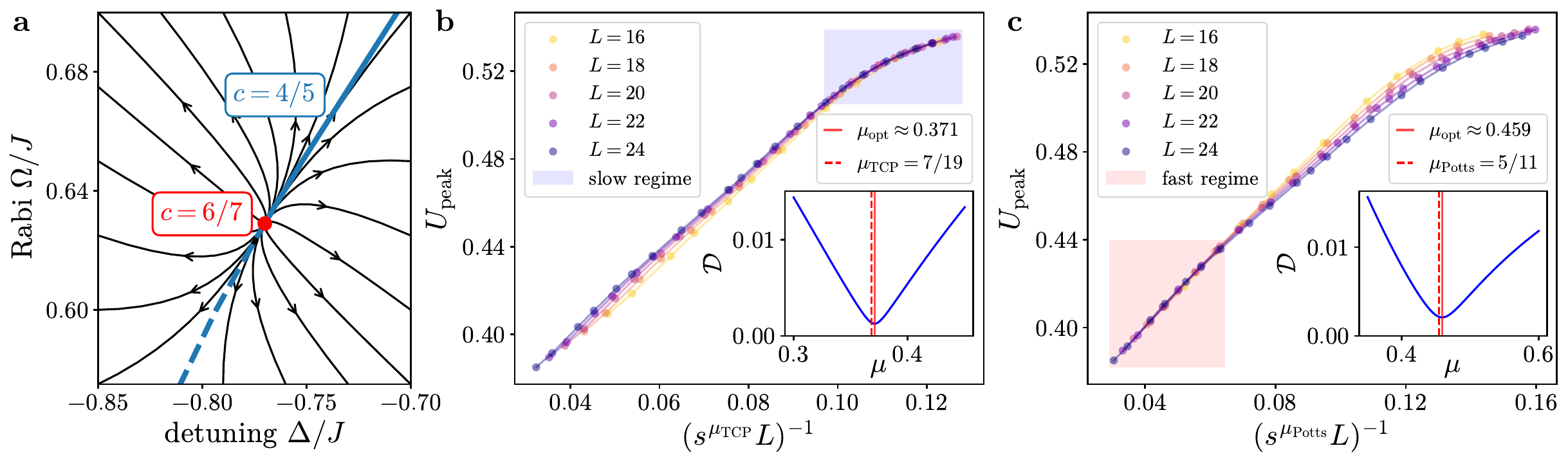}
    \caption{\textbf{a}, RG flow near the TCP. \textbf{b}, Slow ramping data collapse with $\mu_\mathrm{TCP}$. \textbf{c}, Fast ramping data collapse with $\mu_\mathrm{Potts}$.} 
    \label{Fig:TCP_Potts}
        \end{center}
\end{figure*}

\begin{figure*}[t!]
    \begin{center}  
    \includegraphics[width=0.75\columnwidth]{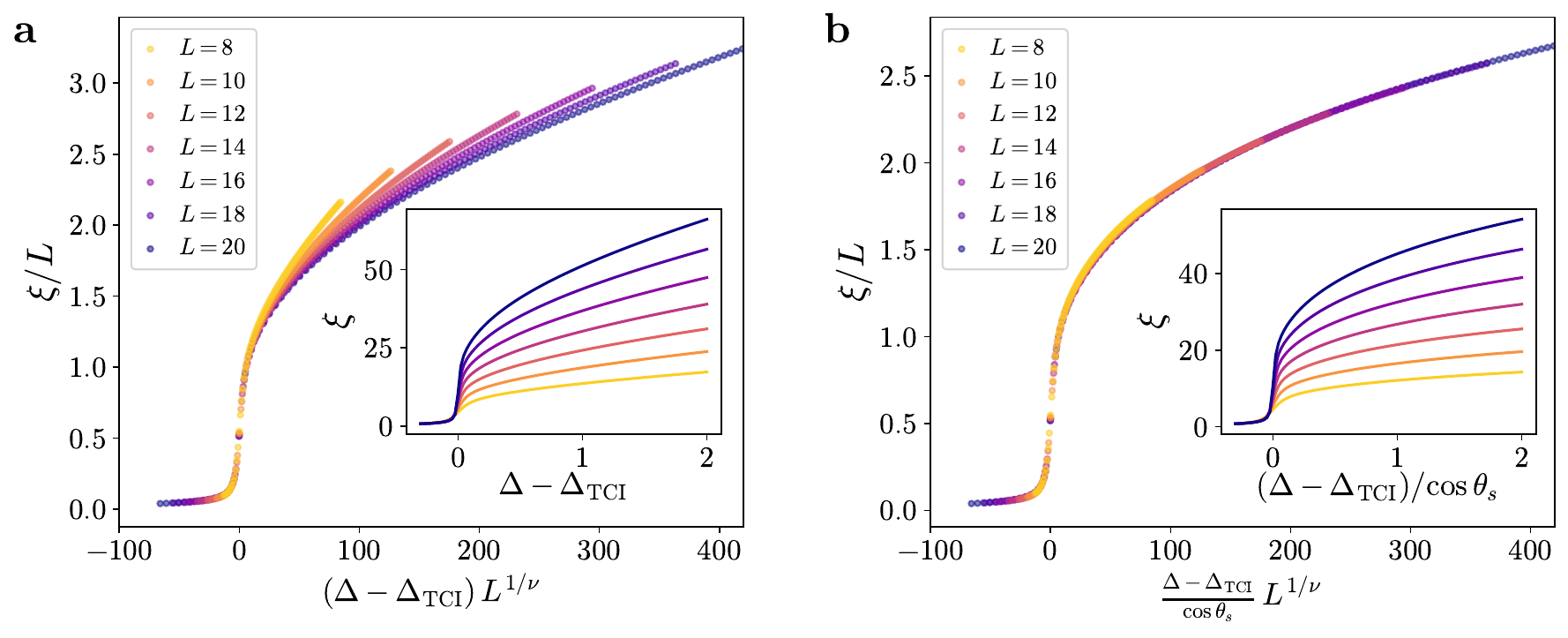}
    \caption{\textbf{a}, When we fix $\Omega=\Omega_\mathrm{TCI}$ and tuning $\Delta$, the data collapse is valid only for a narrow range. \textbf{b}, If we choose the path with $\theta_s \approx 10^\circ$, the data collapse is valid for a much wider range.} 
    \label{fig:scaling}
        \end{center}
\end{figure*}

\end{onecolumngrid}
\end{document}